\documentclass[11pt]{article}
\usepackage{amsfonts,amsmath,amssymb,graphics,epsfig}
\begin{document}

\date{\empty}

\author{D. Kranas$^{1}$\thanks{Current address: Dept of Physics \& Astronomy, Louisiana State University, Baton Rouge, LA 70803, USA.}, C.G. Tsagas$^{1,2}$, John D. Barrow$^{2}$ and D. Iosifidis$^{3}$\\ {\small $^1$Section of Astrophysics, Astronomy and Mechanics, Department of Physics}\\ {\small Aristotle University of Thessaloniki, Thessaloniki 54124, Greece}\\ {\small $^2$DAMTP, Centre for Mathematical Sciences, University of Cambridge}\\ {\small Wilberforce Road, Cambridge CB3 0WA, UK}\\ {\small $^3$Institute of Theoretical Physics, Department of Physics}\\ {\small Aristotle University of Thessaloniki, Thessaloniki 54124, Greece}}

\title{\textbf{Friedmann-like universes with torsion}}

\maketitle

\begin{abstract}
We consider spatially homogeneous and isotropic cosmologies with non-zero torsion. Given the high symmetry of these universes, we adopt a specific form for the torsion tensor that preserves the homogeneity and isotropy of the spatial surfaces. Employing both covariant and metric-based techniques, we derive the torsional versions of the continuity, the Friedmann and the Raychaudhuri equations. These formulae demonstrate how, by playing the role of the spatial curvature, or that of the cosmological constant, torsion can drastically change the evolution of the classic homogeneous and isotropic Friedmann universes. In particular, torsion alone can lead to exponential expansion. For instance, in the presence of torsion, the Milne and the Einstein-de Sitter universes evolve like the de Sitter model. We also show that, by changing the expansion rate of the early universe, torsion can affect the primordial nucleosynthesis of helium-4. We use this sensitivity to impose strong cosmological bounds on the relative strength of the associated torsion field, requiring that its ratio to the Hubble expansion rate lies in the narrow interval ($-0.005813,\,+0.019370$) around zero. Interestingly, the introduction of torsion can \textit{reduce} the production of primordial helium-4, unlike other changes to the standard thermal history of an isotropic universe. Finally, turning to static spacetimes, we find that there exist torsional analogues of the classic Einstein static universe, with all three types of spatial geometry. These models can be stable when the torsion field and the universe's spatial curvature have the appropriate profiles.
\end{abstract}

\section{Introduction}\label{sI}
The Einstein-Cartan (EC) theory is an extension of general relativity that accounts for the presence of spacetime torsion. The theory was first introduced by Cartan in 1922, in order to propose torsion as the macroscopic manifestation of the intrinsic angular momentum (spin) of matter~\cite{C}. Nevertheless, Cartan's suggestion preceded the discovery of quantum spin and his theory did not receive much attention at the time. Many years later, in the 1960s, the spin of the matter was independently reintroduced to general relativity by Kibble and Sciama~\cite{K}. Since then, the EC theory (also known as ECKS theory) has been formally established and has received considerable recognition, as it provides the simplest classical extension of Einstein's general relativity (see~\cite{BH} for a recent review and references therein).

The EC theory postulates an asymmetric affine connection for the spacetime, in contrast to the symmetric Christoffel symbols of Riemannian spaces. In technical terms, torsion is described by the antisymmetric part of the non-Riemannian affine connection~\cite{C}. Therefore, in addition to the metric tensor, there is an independent torsional field, which also contributes to the total gravitational \textquotedblleft pull\textquotedblright. Geometrically speaking, curvature reflects the fact that the parallel transport of a vector along a closed loop in a Riemannian space depends on the path. The presence of torsion adds extra complications, since the aforementioned loop does not necessarily close. In a sense, curvature forces the spacetime to bend and torsion twists it. Dynamically, spacetime torsion is triggered by the intrinsic angular momentum (spin) of the matter, whereas spacetime curvature is caused by the mere presence of matter. This distinction is reflected in two sets of formulae, known as the Einstein-Cartan and the Cartan field equations.

The literature contains a number of suggestions for experimentally testing gravitational theories with non-zero torsion (see~\cite{H} for a representative though incomplete list). As yet, however, there is no experimental or observational evidence to support the distinctive predictions of the EC theory, or the existence of spacetime torsion. The main reason is that the theory only deviates from classical general relativity at extremely high energy densities. These densities can be achieved only in the deep interior of compact objects, like neutron stars and black holes, or during the very early stages of the universe's expansion. Such environments are far beyond our current experimental capabilities. Nevertheless, the EC theory could still provide some answers to the unresolved questions of modern theoretical physics and astrophysics, such as the singularity question, inflationary models, the recently discovered universal acceleration, and what is required of any \textquotedblleft marriage\textquotedblright\ between general relativity and quantum mechanics.

We will investigate spatially homogeneous and isotropic spacetimes with non-zero torsion, which contain matter that may or may not have spin. In other words, torsion can be directly related to the material component, but it might also be treated as an intrinsic property of the host spacetime (see~\cite{CLS} for a related discussion). Given the high symmetry of the latter, the allowed torsion field has to satisfy certain constraints. We therefore adopt a specific form for the torsion tensor that preserves both the spatial homogeneity and the spatial isotropy of the host. This was first introduced in~\cite{T} and falls into the class of the so-called vectorial torsion fields~\cite{O}. Practically speaking, torsion is fully determined by a scalar function that depends only on time. Then, using the Cartan field equations, one can obtain the corresponding expression for the spin. These choices allow us to construct and study the torsional analogues of the standard Friedmann-Robertson-Walker (FRW) universes. In the process we show that, despite the presence of torsion, the high symmetry of the FRW background also ensures the symmetry of the associated Ricci curvature tensor. This in turn implies that the corresponding Einstein and energy-momentum tensors are symmetric as well. Moreover, starting from first principles and using both 1+3 covariant and metric-based techniques, we present the three key formulae monitoring the evolution of these models, namely the Friedmann, the Raychaudhuri, and the continuity equations.\footnote{For the benefit of the readers we provide two appendices, with all the auxiliary relations and the technical information necessary to reproduce our results, at the end of the manuscript.} These enable us to ``quantify'' the torsion input to the total effective energy density of the system, by means of an associated $\Omega$-parameter, as well as its contribution to the deceleration parameter. We also introduce a set of dimensionless variables that measure the relative strength of the torsion effect.

Our solutions are indicative of the occasionally unexpected way torsion can modify the standard evolution of the classic Friedmannian cosmologies. This stems from the fact that torsion can generally play the role of the spatial curvature, as well as reproduce the effects of a cosmological constant, or even those of dark energy. As a result, torsional cosmologies with or without matter can experience accelerated expansion. We find, in particular, that torsion can force the Milne and the Einstein-de Sitter universes into a regime of accelerated expansion analogous to that of their de Sitter counterpart. These examples seem to suggest that a torsion-dominated early universe, or a dust-dominated late-time cosmos, could go through a phase of accelerated expansion without the need of a cosmological constant, the inflaton field, or dark energy. Analogous effects were reported in~\cite{P}, which indicates that the implications of torsion for cosmology deserve further scrutiny.

In an attempt to look for observational signatures of torsion, we find that the latter can affect the outcome of primordial nucleosynthesis, since it changes the expansion rate of the universe. This can be subsequently used to set observational constraints on the allowed torsion fields. Here, assuming that the relative torsion contribution to the expansion remains constant in
time, we are able to calculate its effect on the residual amount of helium-4 produced during primordial nucleosynthesis. Combining this result with the currently allowed range of the primordial helium-4 abundance, leads to a very strong constraint on the strength of the associated torsion field.

Finally, we turn our attention to static spacetimes with non-zero torsion. In particular, we study the structure of the torsional analogue of the Einstein-static universe and also investigate its linear stability. We now find that torsion cannot replace the cosmological constant but it can play the role of the 3-curvature. As a result, there can be static models with non-zero torsion and all three types of spatial geometry, that is Euclidean, spherical or hyperbolic. Our last step is to use standard perturbative techniques to test the linear stability of these new static spacetimes. We find that static solutions with positive curvature seem to be always unstable, while those with zero or negative 3-curvature can achieve stable configurations.

\section{Spacetimes with torsion}\label{sST} 
Riemannian geometry demands the symmetry of the affine connection, thus ensuring torsion-free spaces. Nevertheless, by treating torsion as an independent geometrical field, in addition to the metric, one extends the possibilities to the so-called Riemann-Cartan spaces.

\subsection{Torsion and contortion}\label{ssTC} 
In a general metric space the torsion tensor is defined by the antisymmetric component of the affine connection, namely by $S^{a}{}_{bc}=\Gamma^{a}{}_{[bc]}$ (with $S^{a}{}_{bc}= S^{a}{}_{[bc]}$ being the torsion tensor). Imposing the familiar metricity condition, that is demanding that the metric tensor is covariantly constant (i.e.~$\nabla_{c}g_{ab}=0$), leads to the following decomposition of the generalised (asymmetric) connection
\begin{equation}
\Gamma^{a}{}_{bc}= \tilde{\Gamma}^{a}{}_{bc}+ K^{a}{}_{bc}\,.  \label{Gamma}
\end{equation}
Here, $\tilde{\Gamma}^{a}{}_{bc}$ defines the Christoffel symbols and $K^{a}{}_{bc}$ is the contortion tensor given by\footnote{Alternatively, one may define the torsion and the contortion tensors as $S_{bc}{}^{a}=\Gamma^{a}{}_{[bc]}$ and $K_{ab}{}^{c}=S_{ab}{}^{c}-S_{b}{}^{c}{}_{a}+S^{c}{}_{ab}$, respectively (e.g.~see~\cite{T,S}). In this study, we have adopted the definitions and the conventions of~\cite{PTB}, which follow those of~\cite{P}, though in the latter papers the metric signature is ($+,-,-,-$). Also note that the tildas will always indicate purely Riemannian (torsion free) variables.}
\begin{equation}
K_{abc}= S_{abc}+ S_{bca}+ S_{cba}= S_{abc}+ 2S_{(bc)a}\,, \label{cont}
\end{equation}
with $K_{abc}=K_{[ab]c}$. Seen from the geometrical point of view, torsion prevents infinitesimal parallelograms from closing (e.g.~see~\cite{HvdHK}). Physically speaking, torsion can be seen as a possible link between the intrinsic angular momentum (i.e.~the spin) of the matter and the geometry of the host spacetime. It should also be noted that definitions (\ref{Gamma}) and (\ref{cont}) ensure that $\Gamma^{a}{}_{(bc)}=\tilde{\Gamma}^{a}{}_{bc}+
2S_{(bc)}{}^{a}\neq \tilde{\Gamma}^{a}{}_{bc}$, which means that the
symmetric part of the generalised connection does not necessarily coincide with the Christoffel symbols.

The antisymmetry of the torsion tensor guarantees that it has only one non-trivial contraction, leading to the torsion vector
\begin{equation}
S_a= S^b{}_{ab}= -S^b{}_{ba}\,.  \label{torsv}
\end{equation}
As we will see later, the torsion vector becomes the sole carrier of the torsion effects in spatially homogeneous and isotropic spacetimes. Following (\ref{cont}), there is only one independent contraction of the contortion tensor as well. In particular, we have $K^b{}_{ab}=2S_a=-K_{ab}{}^b$ with $K^b{}_{ba}=0$.

\subsection{Field equations and Bianchi identities}\label{ssFEBIs}
In spacetimes with non-zero torsion, matter and curvature are coupled together by means of the Einstein-Cartan field equations, namely
\begin{equation}
R_{ab}- {\frac{1}{2}}\,Rg_{ab}= \kappa T_{ab}- \Lambda g_{ab}\,,
\label{E-CFEs}
\end{equation}
where $R_{ab}$ is the Ricci tensor, $R=R^{a}{}_{a}$ is the associated scalar and $T_{ab}$ is the energy-momentum tensor of the matter. Note that $R=4\Lambda-\kappa T$, where $T=T^{a}{}_{a}$ and $\kappa=8\pi G$. Although expression (\ref{E-CFEs}) is formalistically identical to its general relativistic counterpart, here both $R_{ab}$ and $T_{ab}$ are generally asymmetric (i.e.~$R_{[ab]}\neq 0$ and $T_{[ab]}\neq 0$) due to the presence
of torsion. The latter is typically coupled to the spin of the matter via the Cartan field equations
\begin{equation}
S_{abc}= -{\frac{1}{4}}\,\kappa
\left(2s_{bca}+g_{ca}s_{b}-g_{ab}s_{c}\right)\,,  \label{CFEs}
\end{equation}
with $s_{abc}=s_{[ab]c}$ and $s_{a}=s^{b}{}_{ab}$ representing the spin tensor and the spin vector of the matter respectively. The trace of (\ref{CFEs}) gives $S_a=\kappa s_a/4$, relating the torsion and the spin vectors directly.

In the presence of torsion, the Bianchi identities acquire a non-zero right-hand side, when compared to their Riemannian analogues. More specifically, we have
\begin{equation}
\nabla_{[e}R^{ab}{}_{cd]}= 2R^{ab}{}_{f[e}S^{f}{}_{cd]} \label{Bian1}
\end{equation}
and
\begin{equation}
R^{a}{}_{[bcd]}= -2\nabla_{[b}S^{a}{}_{cd]}+ 4S^{a}{}_{e[b}S^{e}{}_{cd]}\,,  \label{Bian2}
\end{equation}
where $R_{abcd}$ is the curvature tensor. Due to non-zero torsion, the latter does not generally satisfy all the symmetries of its Riemannian counterpart (i.e.~$R_{abcd}=R_{[ab][cd]}$ only). Contracting the above given Bianchi identities twice, we arrive at
\begin{equation}
\nabla^{b}G_{ba}= 2R_{bc}S^{cb}{}_{a}+ R_{bcda}S^{dcb}  \label{GD1}
\end{equation}
and
\begin{equation}
G_{[ab]}= 2\nabla_{[a}S_{b]}+ \nabla^{c}S_{cab}- 2S^{c}S_{cab}\,,
\label{Gant1}
\end{equation}
respectively. Note that $G_{ab}=R_{ab}-(R/2)g_{ab}$ defines the torsional analogue of the familiar Einstein tensor, which is generally asymmetric as well (i.e.~$G_{[ab]}\neq0$ -- see Eq.~(\ref{Gant1})). Finally, relation (\ref{Gant1}) gives
\begin{equation}
\nabla^{b}G_{ab}= \nabla^{b}G_{ba}-
2\left(\nabla^{2}S_{a}-\nabla^{b}\nabla_{a}S_{b}
+\nabla^{b}\nabla^{c}S_{cba}\right)- 4\nabla^{b}\left(S^{c}S_{cab}\right)\,,  \label{NbGab}
\end{equation}
ensuring that the general relativistic conservation law $\nabla^{b}G_{ab}=0$ does not generally hold in the presence of spacetime torsion.

\subsection{Kinematics}\label{ssKs} 
Introducing the timelike 4-velocity field $u_{a}$ (with $u_{a}u^{a}=-1$) leads to the 1+3 decomposition of the spacetime into time and 3-dimensional space (see~\cite{TCM} for an extended review of the formalism and its applications). In particular, the metric tensor splits as $g_{ab}=h_{ab}-u_{a}u_{b}$, where $h_{ab}$ is a symmetric 3-tensor orthogonal to $u_{a}$ (i.e.~$h_{ab}= h_{(ab)}$, $h_{ab}u^{b}=0$ and $h_{a}{}^{a}=3$), known as the projection tensor. In the absence of rotation, the latter also acts as the metric tensor of the 3-dimensional spatial hypersurfaces. The
kinematics of the aforementioned 4-velocity field are decoded by splitting its covariant gradient as
\begin{equation}
\nabla_{b}u_{a}= {\frac{1}{3}}\,\Theta h_{ab}+ \sigma_{ab}+ \omega_{ab}- A_{a}u_{b}\,.  \label{Nbua}
\end{equation}
Here, $\Theta=\mathrm{D}^{a}u_{a}$ is the volume scalar, $\sigma_{ab}=\mathrm{D}_{\langle b}u_{a\rangle}$ and $\omega_{ab}= \mathrm{D}_{[b}u_{a]}$ are respectively the shear and the vorticity tensors, while $A_{a}=\dot{u}_{a}$ is the 4-acceleration vector.\footnote{Overdots indicate temporal derivatives (along the timelike $u_{a}$-field). For instance $A_{a}=\dot{u}_{a}= u^{b}\nabla_{b}u_{a}$ by definition. Spatial derivatives (orthogonal to $u_{a}$), on the other hand, are denoted by the covariant operator $\mathrm{D}_{a}=h_{a}{}^{b}\nabla_{b}$. Therefore, $\Theta =\mathrm{D}^{a}u_{a}=h^{ab}\nabla _{b}u_{a}$, $\sigma_{ab}= \mathrm{D}_{\langle b}u_{a\rangle}=h_{\langle b}{}^{d} h_{a\rangle}{}^{c}\nabla_{d}u_{c}$, etc (see~\cite{TCM} for more details). Also, round brackets denote symmetrisation and square antisymmetrisation, while angled ones indicate the symmetric and trace-free part of second rank tensors (e.g.~$\sigma_{ab}= \mathrm{D}_{\langle b}u_{a\rangle}=\mathrm{D}_{(b}u_{a)}- (\mathrm{D}^{c}u_{c}/3)h_{ab}$ by construction).} The volume scalar monitors the convergence/divergence of the worldlines tangent to
4-velocity field, while the shear and the vorticity tensors describe
kinematic anisotropies and the rotational behaviour of the $u_{a}$-field respectively. Finally, a non-zero 4-acceleration vector implies that the aforementioned worldlines are not autoparallel curves (see~\cite{PTB} for a discussion on the distinction between geodesics and autoparallel curves in spaces with non-zero torsion).

The rate of convergence/divergence of a worldline congruence is governed by the Raychaudhuri equation. In the presence of torsion, the latter reads~\cite{PTB}
\begin{eqnarray}
\dot{\Theta}&=& -{\frac{1}{3}}\,\Theta^{2}- R_{(ab)}u^{a}u^{b}-
2\left(\sigma^{2}-\omega^{2}\right)+ \mathrm{D}_{a}A^{a}+ A_{a}A^{a}  \notag\\ &&+{\frac{2}{3}}\,\Theta S_{a}u^{a}- 2S_{(ab)c}u^{a}u^{b}A^{c}- 2S_{\langle ab\rangle c}\sigma^{ab}u^{c}+ 2S_{[ab]c}\omega^{ab}u^{c}\,,  \label{Ray1}
\end{eqnarray}
where only the symmetric component of the (generally asymmetric) Ricci tensor contributes to the right-hand side. Analogous propagation formulae can be obtained for the shear and the vorticity. These are supplemented by a set of three constraint equations, relating the gradients of the kinematic variables. Here, however, we will focus on the mean spacetime kinematics, that is on its contraction or expansion, and refer the interested reader to~\cite{PTB} for further discussion.

\section{FRW-like models with torsion}\label{sDRW-TMT} 
A spatially homogeneous and isotropic, Friedmann-like, spacetime cannot naturally accommodate an arbitrary form of torsion. In what follows, we will investigate the implications of such highly symmetric torsion fields for the evolution of the cosmological spacetime.

\subsection{The torsion field}\label{ssTA} 
Consider an FRW-type spacetime with non-zero torsion and a family of observers living along a timelike congruence tangent to the 4-velocity field $u_{a}$, (as defined in the previous section). In order to preserve the homogeneity and isotropy of a maximally symmetric 3-dimensional rest-space of these observers, we adopt the following form for the torsion tensor~\cite{T}
\begin{equation}
S_{abc}= 2\phi\,h_{a[b}u_{c]}\,.  \label{ans.ten}
\end{equation}
Note that $\phi$ is a scalar function that depends only on time (i.e.~$\phi=\phi(t)$), since any spatial dependence is forbidden by the homogeneity of the 3-space. The above given torsion field also respects isotropy, which becomes clearer after evaluating the associated torsion vector. Indeed, contracting (\ref{ans.ten}) and using definition (\ref{torsv}), leads to
\begin{equation}
S_{a}= -3\phi u_{a}\,, \hspace{5mm} \mathrm{so~that} \hspace{5mm}
\begin{cases}
\phi>0\,, \hspace{3mm} \mathrm{when} \hspace{3mm} S_{a}\downarrow\uparrow u_{a}\,, &  \\ \phi<0\,, \hspace{3mm} \mathrm{when} \hspace{3mm} S_{a}\uparrow\uparrow u_{a}\,. &
\end{cases}  \label{ans.vec}
\end{equation}
Not surprisingly, the associated torsion vector is timelike (see also~\cite{PTB}), in agreement with the isotropy of the spatial sections.\footnote{On using the Cartan field equations (see (\ref{CFEs}) in \S ~\ref{ssFEBIs}), one can recast (\ref{ans.ten}) and (\ref{ans.vec}) into the expressions $\kappa s_{abc}=8\phi h_{c[a}u_{b]}$ and $\kappa s_a=12\phi u_a$ for the spin tensor and the spin vector respectively. These in turn combine with Eqs.~(\ref{ans.ten}) and (\ref{ans.vec}) to guarantee that the torsion and the spin fields are related by
\begin{equation}
S_{abc}= -{\frac{1}{4}}\,\kappa s_{cba} \hspace{10mm} \mathrm{and} \hspace{10mm} S_a= -{\frac{1}{4}}\,\kappa s_a\,.  \label{TvsS}
\end{equation}
One could use the above to replace torsion with spin in our formulae. Nevertheless, given that the two fields are simply proportional to each other, we will proceed with our study focusing on torsion rather than spin.} It is also worth noting that, according to Eq.~(\ref{ans.vec}), the sign of $\phi$ is directly connected to the relative orientation between the torsion and the 4-velocity vectors. In particular, for negative values of $\phi$ the torsion vector is future-directed, while in the opposite case $S_{a}$ becomes past-directed. Finally, applying (\ref{ans.ten}) to definition (\ref{cont}), we find that
\begin{equation}
K_{abc}= 4\phi\,u_{[a}h_{b]c}\,,  \label{FRWcont}
\end{equation}
which provides the contortion tensor in FRW-like spacetimes with torsion. The above implies that $K^{b}{}_{ab}=-6\phi u_{a}=2S_{a}= -K_{ab}{}^{b}$ and $K^{b}{}_{ba}=0$ as expected (see \S~\ref{ssTC} previously).

\subsection{Conservation laws}\label{ssCLs} 
Applying (\ref{ans.ten}) and (\ref{ans.vec}) to the second of the
twice-contracted Bianchi identities (see Eq.~(\ref{Gant1}) in \S~\ref{ssFEBIs}), it is straightforward to show that the right-hand side of the latter relation vanishes. This ensures that $G_{[ab]}=0$, which in turn guarantees that $R_{[ab]}=0$ and $T_{[ab]}=0$ as well. Consequently, in spatially homogeneous and isotropic spacetimes, the Ricci and the energy-momentum tensors retain their familiar (Riemannian) symmetry despite the presence of torsion.\footnote{When showing the symmetry of the Ricci and energy-momentum tensors in FRW-like models, one also needs to account for the fact that 4-velocity decomposition (see Eq.~(\ref{Nbua}) in \S~\ref{ssKs}) reduces to $\nabla_{b}u_{a}= (\Theta /3)h_{ab}$ and that $\nabla_{a}\phi=-\dot{\phi}u_{a}$ (since $\mathrm{D}_{a}\phi=0$ by default) in these highly symmetric spacetimes.}

Introducing our form of torsion to the Einstein-Cartan field equations and using the first of the twice-contracted Bianchi identities (see Eqs.~(\ref{E-CFEs}) and (\ref{GD1}) respectively), leads to the constraint
\begin{equation}
\nabla^{b}T_{ab}= -4\phi\left(T_{ab}u^{b}-\kappa^{-1}\Lambda u_{a}\right)\,,  \label{cons.law}
\end{equation}
with the right-hand side vanishing in the absence of torsion. Moreover, given the high symmetry of the host spacetime, the matter must have the form of a perfect fluid with
\begin{equation}
T_{ab}= \rho u_{a}u_{b}+ ph_{ab}\,,  \label{emt1}
\end{equation}
where $\rho$ and $p$ represent its energy density and isotropic pressure respectively (we assume no bulk viscosity). Substituting the above into (\ref{cons.law}), one obtains the continuity equation
\begin{equation}
\dot{\rho}= -\Theta\left(\rho+p\right)+
4\phi\left(\rho+\kappa^{-1}\Lambda\right)\,.  \label{cont1}
\end{equation}
The latter provides the conservation law of the matter energy density in Friedmann-type cosmologies with non-zero torsion. It is also straightforward to verify that, for vanishing torsion (i.e.~when $\phi=0$), Eq.~(\ref{cont1}) recovers its familiar general relativistic expression.

\subsection{The Raychaudhuri equation}\label{ssRE} 
Demanding spatial homogeneity and isotropy implies $\sigma_{ab}=0=
\omega_{ab}=A_{a}$. Then, applying (\ref{ans.ten}) and (\ref{ans.vec}) to the generalised Raychaudhuri equation (see (\ref{Ray1})), the latter simplifies to
\begin{equation}
\dot{\Theta}= -{\frac{1}{3}}\,\Theta^{2}- {\frac{1}{2}}\,\kappa\left(\rho+3p\right)+ \Lambda+ 2\Theta\phi\,.  \label{FRWRay1}
\end{equation}
Note that in deriving the above we have also used the Einstein-Cartan equations, assuming that matter has the form of a perfect fluid. The last term on the right-hand side of Eq.~(\ref{FRWRay1}) implies that torsion assists or inhibits the expansion/contraction of the timelike congruence (i.e.~the one tangent to the $u_{a}$-field), depending on the sign of $\phi$. The latter in turn depends on whether the torsion vector is future-directed or past-directed (see \S~\ref{ssTA} before).

The volume scalar ($\Theta$) of a spacetime with non-zero torsion and its purely Riemannian (i.e.~torsion-free) counterpart ($\tilde{\Theta}$) are related by~\cite{PTB}
\begin{equation}
\Theta= \tilde{\Theta}+ K^{a}{}_{ba}u^{b}\,.  \label{Thetas1}
\end{equation}
Recalling that the contortion tensor of an FRW-type model with torsion satisfies constraint (\ref{FRWcont}), the above relation reduces to
\begin{equation}
\Theta= \tilde{\Theta}+ 6\phi= 3\left({\frac{\dot{a}}{a}}\right)+ 6\phi= 3H\left(1+2\,{\frac{\phi}{H}}\right)\,,  \label{Thetas2}
\end{equation}
given that $\dot{a}/a=\tilde{\Theta}/3=H$ defines the cosmological scale factor ($a=a(t)$) both in torsional and in torsion-free Friedmannian cosmologies, with $H$ being the associated Hubble parameter.\footnote{Following (\ref{Thetas2}), the dimensionless ratio $\phi/H$ measures the \textquotedblleft relative strength\textquotedblright of the torsion effects.} According to (\ref{Thetas2}), the divergence/convergence of a worldline congruence in an FRW-type cosmology with torsion is not solely determined by the scale-factor evolution. Substituting relation (\ref{Thetas2}) back into Eq.~(\ref{FRWRay1}), we obtain
\begin{equation}
{\frac{\ddot{a}}{a}}= -{\frac{1}{6}}\,\kappa\left(\rho+3p\right)+ {\frac{1}{3}}\,\Lambda- 2\dot{\phi}- 2\left({\frac{\dot{a}}{a}}\right)\phi\,.  \label{FRWRay2}
\end{equation}
The latter provides an alternative expression of the Raychaudhuri equation in FRW-type cosmologies with non-zero torsion, this time in terms of the model's scale factor.

\subsection{The Friedmann equations}\label{ssFEs} 
Treating the torsion field independently of the metric, means that the line element of the host spacetime is identical to its Riemannian counterpart. Therefore, in the case of an FRW-type cosmology with torsion, we have
\begin{equation}
\mathrm{d}s^{2}= -\mathrm{d}t^{2}+ a^{2}\left[(1-Kr^{2})^{-1}\mathrm{d}r^{2} +r^{2}\mathrm{d}\vartheta^{2} +r^{2}\sin^{2}\vartheta\mathrm{d}\varphi\right]\,,  \label{FRWle}
\end{equation}
where $a=a(t)$ is the scale factor (with $\dot{a}/a= \tilde{\Theta}/3=H$ -- see also relation (\ref{Thetas2}) in the previous section) and $K=0,\pm1$ is the 3-curvature index. The above metric, together with Eq.~(\ref{ans.ten}) and the metricity condition (i.e.~$\nabla_{c}g_{ab}=0$), provides the components of the generalized connection (see Appendix~A for the details). One can then evaluate the components of the Ricci tensor, and subsequently the Ricci scalar, by recalling that
\begin{equation}
R_{ab}= -\partial_{b}\Gamma^{c}{}_{ac}+ \partial_{c}\Gamma^{c}{}_{ab}- \Gamma^{e}{}_{ac}\Gamma^{c}{}_{eb}+ \Gamma^{e}{}_{ab}\Gamma^{c}{}_{ec}  \label{Ricci}
\end{equation}
and then employing a lengthy calculation (see Appendix~B). Finally, assuming a perfect fluid and involving the Einstein-Cartan field equations, we arrive at
\begin{equation}
\left({\frac{\dot{a}}{a}}\right)^{2}= {\frac{1}{3}}\,\kappa\rho- \frac{K}{a^{2}}+ {\frac{1}{3}}\,\Lambda- 4\phi^{2}- 4\left({\frac{\dot{a}}{a}}\right)\phi  \label{Fried1}
\end{equation}
and
\begin{equation}
{\frac{\ddot{a}}{a}}= -{\frac{1}{6}}\,\kappa\left(\rho+3p\right)+ {\frac{1}{3}}\,\Lambda- 2\dot{\phi}- 2\left({\frac{\dot{a}}{a}}\right)\phi\,.  \label{Fried2}
\end{equation}
These are the torsional analogues of the Friedmann equations, obtained here by means of metric-based techniques.\footnote{Recalling that $\Theta=3H=3\dot{a}/a+6\phi$ (see relation (\ref{Thetas2}) in \S~\ref{ssRE}) and then using it to eliminate the scale factor from Eq.~(\ref{Fried1}), the latter assumes the covariant form
\begin{equation}
{\frac{1}{9}}\,\Theta^{2}= {\frac{1}{3}}\,\kappa\rho- {\frac{K}{a^{2}}}+ {\frac{1}{3}}\,\Lambda\,,  \label{covFried1}
\end{equation}
which is formalistically identical to that of its torsion-free counterpart (e.g.~see~\cite{TCM}).} Note that the last expression, namely the Raychaudhuri equation of an FRW-type cosmology with torsion, is identical to the one derived earlier by employing covariant methods (compare to Eq.~(\ref{FRWRay2}) in \S~\ref{ssRE}). This verifies the consistency of our analysis. Additional agreement comes by showing that the continuity equation obtained from Eqs.~(\ref{Fried1}) and (\ref{Fried2}) is identical to the one derived earlier (see relation (\ref{cont1}) in \S~\ref{ssCLs}).

Following Eq.~(\ref{Fried1}), torsion contributes to the total effective energy-density of the system. More specifically, the torsional analogue of the Friedmann equation recasts as
\begin{equation}
1= \Omega_{\rho}+\Omega_K+\Omega_{\Lambda}+ \Omega_{\phi}\,,  \label{Omegas}
\end{equation}
where $\Omega_{\rho}=\kappa\rho/3H^2$, $\Omega_K=-K/a^2H^2$, $\Omega_{\Lambda}=\Lambda/3H^2$ and $\Omega_{\phi}= -4[1+(\phi/H)](\phi/H)$ are the associated density parameters. The strength of the torsion contribution, relative to that of the matter for example, is measured by the dimensionless ratio $\Omega_{\phi}/\Omega_{\rho}$. Following (\ref{Omegas}), the torsion contribution to the Friedmann equation vanishes when $\phi/H=0,-1$. On the other hand, torsion dominates completely when $\phi/H=-1/2$, which translates into $\Omega_{\phi}\simeq1$ and vice versa. In an expanding universe (where $H>0$), the latter can occur only for $\phi<0$ (see also Eq.~(\ref{Fried1}) earlier).

Starting from (\ref{Fried2}) and keeping in mind that $q=-\ddot{a}a/\dot{a}^2 $ defines the deceleration parameter of the universe, we may write
\begin{equation}
qH^2= {\frac{1}{6}}\,\kappa(\rho+3p)- {\frac{1}{3}}\,\Lambda+ 2\dot{\phi}+ 2H\phi\,,  \label{tq}
\end{equation}
The above implies that the torsion field can either assist or inhibit accelerated expansion (that with $q<0$). When $\phi$ is constant and negative, in particular, the presence of torsion tends to accelerate the expansion (see also \S~\ref{sssEVT-DSs} and \S~\ref{sssESM} below).

\subsection{Characteristic solutions}\label{ssCSs} 
Assuming zero cosmological constant and using relation (\ref{Thetas2}), the continuity equation (see expression (\ref{cont1}) in \S~\ref{ssCLs}) of a barotropic medium with $p=w\rho$, reads
\begin{equation}
{\frac{\dot{\rho}}{\rho}}= -3(1+w)\left({\frac{\dot{a}}{a}}\right)-
2(1+3w)\phi\,.  \label{bcont}
\end{equation}
When $w=$~constant the above integrates to
\begin{equation}
\rho= \rho_0\left({\frac{a}{a_0}}\right)^{-3(1+w)} \mathrm{exp}\left[-2(1+3w)\int_{t_0}^t\phi\mathrm{d}t\right]\,,  \label{brho1}
\end{equation}
with $\rho_0=\rho(t=t_0)$ and $a_0=a(t=t_0)$. Accordingly, the torsioneffect on the energy-density evolution (propagating via the exponential term on the right-hand side of the above) generally depends on the equation of state of the matter. The torsion contribution to the right-hand side of (\ref{brho1}) vanishes in the special case of a medium with zero effective gravitational mass/energy (i.e.~for $w=-1/3$). On the other hand, the energy-density evolution becomes essentially torsion dominated in the case of a vacuum stress with $w=-1$. Then, allowing also for non-zero cosmological constant, Eq.~(\ref{bcont}) gives
\begin{equation}
\rho= \left(\rho_0+\kappa^{-1}\Lambda\right) \mathrm{exp}\left(4\int_{t_0}^t\phi\mathrm{d}t\right)- \kappa^{-1}\Lambda\,,  \label{brho2}
\end{equation}
which reduces to $\rho= \rho_0\,\mathrm{exp}(4\int_{t_0}^t\phi\mathrm{d}t)$ in the absence of a cosmological constant. Therefore, the energy density generally varies in time due to the presence of torsion, which means that a
fluid with $p=-\rho$ is not dynamically equivalent to a cosmological
constant. Such a behaviour, which is in direct contrast with the purely general relativistic picture, has been encountered in scalar-tensor theories like the Brans-Dicke theory~\cite{N}.

\subsubsection{Exact vacuum and torsion-dominated 
solutions}\label{sssEVT-DSs}
In standard general relativity empty spacetimes with Euclidean spatial sections and no cosmological constant are static. Torsion can drastically change this picture. Indeed, when $\rho=0=K= \Lambda$, the torsional analogue of the Friedmann equation (see expression (\ref{Fried1}) in \S~\ref{ssFEs}) recasts into the perfect square $(\dot{a}/a+2\phi)^2=0$. The latter ensures
the following relation
\begin{equation}
\phi= \phi(t)= -{\frac{1}{2}}\,{\frac{\dot{a}}{a}}\,, \label{vacphi1}
\end{equation}
between the torsion scalar and the cosmological scale factor. Then, the choice $\phi=\phi_0=$~constant leads to $\dot{a}/a=$~constant and subsequently to the de Sitter-like expansion
\begin{equation}
a= a_{0}\mathrm{e}^{-2\phi_{0}(t-t_{0})}\,,  \label{tMilne2}
\end{equation}
when $\phi_0<0$. Note that, to first approximation, the same solution also governs the evolution of a Friedmann-like universe with matter, provided that torsion dominates (i.e.~for $\Omega_{\phi}\gg\Omega_{\rho}$ - see Eq.~(\ref{Omegas}) in \S~\ref{ssFEs}). Therefore, a torsion-dominated early universe could have in principle undergone a phase of inflationary expansion without the need of a cosmological constant, or the presence of an inflaton field.

The picture does not change if we adopt a more general ansatz for the torsion scalar. For instance, setting $\phi=\phi(t)= \phi_0+At^n$ (where $\phi_0$, $A$ and $n\neq-1$ are constants),\footnote{When $n=-1$ and $\phi\propto t^{-1}$, Eq.~(\ref{vacphi1}) integrates to the power-law solution $a\propto t^{-2\phi_0t_0}$ (where $\phi_0= \phi(t=t_0)$).} Eq.~(\ref{vacphi1}) yields
\begin{equation}
a= a_0\mathrm{exp}\left\{-2\left[\phi_0(t-t_0)+{\frac{A}{n+1}}%
\left(t^{n+1}-t_0^{n+1}\right)\right]\right\}\,.  \label{vacsol1}
\end{equation}
This solution, which (approximately) also holds for torsion-dominated Friedmann-like universes with matter, can lead to exponential expansion as well (depending on the values of the parameters $\phi_0$, $A$ and $n$). In fact, when $n<-1$, the second term inside the square brackets depletes with time and the above asymptotically approaches the de Sitter-like solution of (\ref{tMilne2}) at late times (i.e.~when $t\gg t_0$).

Let us now consider a vacuum spacetime with zero cosmological constant, but this time allow for hyperbolic spatial geometry.\footnote{Vacuum torsional spacetimes with no cosmological constant and spherical spatial geometry do not exist in our scheme. Indeed, in such an environment Eq.~(\ref{Fried1}) recasts into $(\dot{a}/a+2\phi)^2=-1/a^2$, which is impossible.} When $\rho=0=p=\Lambda$, $K=-1$ and $\phi\neq0$, we obtain what one might call the torsion analogue of the classical Milne universe. Then,
Eq.~(\ref{Fried1}) factorises as $(\dot{a}/a+2\phi+1/a) (\dot{a}/a+2\phi-1/a)=0$, giving
\begin{equation}
\phi= \phi(t)= -{\frac{1}{2}}\left({\frac{\dot{a}}{a}}\pm{\frac{1}{a}}\right)\,.  \label{vacphi2}
\end{equation}
Setting $\phi=\phi_0=$~constant on the left-hand side, the above integrates to
\begin{equation}
a= a_0\mathrm{e}^{-2\phi_0(t-t_0)}+ {\frac{1}{2}}\,\phi_0^{-1} \left[\mathrm{e}^{-2\phi_0(t-t_0)}-1\right]\,,  \label{tMilne3}
\end{equation}
which implies exponential expansion when $\phi_0<0$. Therefore, instead of obeying the ``coasting'' solution (with $a=a(t)=t$) of its classical counterpart, the torsional Milne universe exhibits a de Sitter-type behaviour. For all practical purposes, torsion has been playing the role of a (positive) cosmological constant. Note that solution (\ref{tMilne3}) holds (approximately) in the presence of matter as well, provided that $\Omega_{\phi}\gg\Omega_{\rho}$.

\subsubsection{Exact solutions with matter}\label{sssESM} 
We will now sift our attention from vacuum models to those containing matter. In so doing, we assume a Friedmann-like universe with zero 3-curvature and no cosmological constant. When dealing with pressure-free matter (i.e.~dust) and torsion with $\phi= \phi_0=$~constant, Eqs.~(\ref{Fried1}) and (\ref{Fried2}) combine to give
\begin{equation}
{\frac{\ddot{a}}{a}}+ {\frac{1}{2}}\left({\frac{\dot{a}}{a}}\right)^2+
4\phi_0\left({\frac{\dot{a}}{a}}\right)+ 2\phi_0^2= 0\,. \label{tFRWdust}
\end{equation}
The above accepts a solution of the form
\begin{equation}
a^{3/2}= \mathcal{C}_1\mathrm{e}^{-\phi_0t}+ \mathcal{C}_2\mathrm{e}^{-3\phi_0t}\,,  \label{tFRWds1}
\end{equation}
with $\mathcal{C}_{1,2}$ being the integration constants. Clearly, when $\phi_0<0$ the scale factor evolves as
\begin{equation}
a= a_0\mathrm{e}^{-2\phi_0(t-t_0)}\,,  \label{tFRWds2}
\end{equation}
at late times. Note that we have made no a priori assumption on the relative strength of the torsion field. Taken at face value, this means that even the mere presence of torsion could drive the Einstein-de Sitter universe into an accelerated regime analogous to that of the de Sitter model. Without further scrutiny, however, it would be rather premature to claim that torsion can provide a viable alternative answer to the question of the recent universal acceleration. Having said that, we should also mention that analogous claims have been made in~\cite{P}.

False-vacuum cosmologies have been typically associated with inflation, being the driving force of the exponential expansion. In what follows, we will consider torsional FRW-like universes with false-vacuum barotropic index $w=-1$ and zero 3-curvature. We will also allow for a non-zero cosmological constant, although its presence does not alter the nature of the solutions. Then, on using (\ref{brho2}), the associated Friedmann equation (see (\ref{Fried1}) in \S~\ref{ssFEs}) recasts into
\begin{equation}
H= -2\phi\pm \sqrt{{\frac{1}{3}}\left(\kappa\rho_0+\Lambda\right)}\; \mathrm{e}^{2\int_{t_0}^t\phi\mathrm{d}t}\,.  \label{fvacH} \end{equation}
Thus, due to the presence of torsion, the Hubble parameter now varies in time, in contrast to the standard torsionless case where $H=H_0=$~constant. Recalling that $H=\dot{a}/a$ and setting $\phi= \phi_0=$~constant, the above integrates to give
\begin{equation}
a= a_0\;\mathrm{exp}\left[2\phi_0(t-t_0)\pm \sqrt{{\frac{\kappa\rho_0+\Lambda}{12\phi_0^2}}}\, \left(\mathrm{e}^{2\phi_0(t-t_0)}-1\right)\right]\,. \label{fvacsol1}
\end{equation}
When $\phi_0<0$, namely for future-directed torsion vector -- see \S~\ref{ssTA} earlier, the above asymptotically reduces to solution (\ref{tMilne2}) at late times. For $\phi_0>0$, on the other hand, the late-time evolution of the cosmological scale factor is monitored by
\begin{equation}
a= a_0\;\mathrm{exp} \left[\pm\sqrt{{\frac{\kappa\rho_0+\Lambda}{12\phi_0^2}}}\, \mathrm{e}^{2\phi_0(t-t_0)}\right]\,.  \label{fvacsol2}
\end{equation}
In either case the models undergo exponential de Sitter-type inflation, analogous to that of their torsion-free counterparts, although (\ref{fvacsol2}) allows for exponential \textquotedblleft
deflation\textquotedblright as well. One can also extract graduated
inflationary solutions (see~\cite{B} for a discussion) by setting $\phi\propto t^{-1}$. Substituted into Eq.~(\ref{Fried1}), this choice leads to
\begin{equation}
a= a_0\left({\frac{t}{t_0}}\right)^{-2\phi_0t_0}\mathrm{exp} \left[\pm\sqrt{{\frac{(\kappa\rho_0+\Lambda)t_0^2}{3(2\phi_0t_0+1)^2}}} \left({\frac{t}{t_0}}\right)^{2\phi_0t_0+1}\right]\,.  \label{fvacsol3}
\end{equation}

The solutions presented in the last two sections are characteristic of the versatile and occasionally surprising nature of the torsion effects, even when the torsion field takes the very restricted form imposed by the high symmetry of the Friedmann-like host.

\section{Observational bounds on cosmic torsion}\label{sOBCT} 
The literature contains a number of proposals for observational tests of torsion, the majority of which work within the realm of our solar system~\cite{H}. Here, we will make an attempt to put cosmological bounds on the torsion field. Generally, tests for non-zero torsion using astronomical objects, or standard solar-system tests, are very weak. Torsion is linked to matter and does not propagate its effects via a wave equation. However, it does gravitate and so its contribution to the expansion dynamics of the universe offers the prospect of a strong observational test.

\subsection{Steady-state torsion}\label{ssSST} 
A measure of the torsion contribution is given by the dimensionless ratio $\phi /H$. In what follows we will consider torsion fields with $\phi\propto H$. Then, assuming Euclidean 3-spaces, no cosmological constant and setting
\begin{equation}
\lambda =\phi /H,  \label{lam}
\end{equation}
where $\lambda$ is constant, the Friedmann and the continuity equations (see (\ref{Fried1}) and (\ref{bcont})) become
\begin{equation}
H^{2}= {\frac{\kappa \rho }{3(2\lambda+1)^{2}}}\,,  \label{lamFried}
\end{equation}
with $\lambda\neq-1/2$, and
\begin{equation}
{\frac{\dot{\rho}}{\rho}}= -[3+2\lambda+3w(1+2\lambda)]H\,,
\label{lamcont}
\end{equation}
respectively. The former relation shows that torsion changes the Hubble-flow rate, which means that it can affect physical interactions that are sensitive to the rate of the cosmic expansion, like primordial nucleosynthesis of helium-4 for example (see \S~\ref{ssPNBT} next). Suppose now \textquotedblleft steady-state\textquotedblright\ torsion with $\lambda=\phi/H=$~constant. In other words, assume that the torsion contribution to the volume expansion (see Eq.~(\ref{Thetas2}) in \S~\ref{ssRE}) does not change in time. Then, given that $H=\dot{a}/a$, the above two relations combine to give the exact solution
\begin{equation}
a= a_{0}\left({\frac{t}{t_{0}}}\right)^{2/[3+2\lambda +3w(1+2\lambda)]}\,,  \label{lama}
\end{equation}
when $\lambda\neq-1/2$. Recall that the value $\lambda=-1/2$ ensures that $\Omega_{\phi }=1$ (see Eq.~(\ref{Omegas}) in \S~\ref{ssFEs}), which corresponds to the purely torsional FRW-like universe examined above in \S~\ref{sssEVT-DSs}. Next, we will employ this solution to impose cosmological bounds on $\lambda$ and on torsion itself. In doing so, we will turn to the early universe and, specifically, to the epoch of primordial nucleosynthesis.

\subsection{Primordial nucleosynthesis bounds on 
torsion}\label{ssPNBT}
During the radiation era of the early universe, when $w=1/3$, the Friedmann solution (\ref{lama}) reduces to $a\propto t^{1/2(1+\lambda)}$, with $\lambda\neq-1$. Note that for $\lambda=0$ we recover the evolution law of the usual radiation-dominated, torsion-free FRW universe. Recall also that, when $\lambda= \phi/H=-1$, the torsion input to the Friedmann equation vanishes (see expression (\ref{Omegas}) in \S~\ref{ssFEs}). The above,
together with Eq.~(\ref{lamFried}), lead to the following expression for the radiation energy density
\begin{equation}
\rho_{(\gamma)}= {\frac{3(1+2\lambda)}{4\kappa(1+\lambda)^{2}t^{2}}}= \sigma T^{4}\,,  \label{tSB}
\end{equation}
where $\sigma$ is the black-body constant. Consequently, due to the presence of $\lambda $, the time evolution of the temperature differs from that of the standard (torsionless) Friedmann universe (with three light neutrino species).

The freeze-out temperature ($T_{fr}$) of the neutron-proton kinetic equilibrium occurs when the weak interaction rate ($\Gamma_{wk}\propto T^{5} $) for the neutron-proton exchanges ($n\leftrightarrow p+e^{-}+\bar{\nu}_{e}$, $n+\nu_{e}\leftrightarrow p+e^{-}$ and $n+e^{+}\leftrightarrow p+\bar{\nu}_{e}$) equals the Hubble expansion rate. This yields a simple analytic expression for the ratio of the freeze-out temperatures between cosmologies with torsion ($T_{fr}\equiv T_{fr}(\lambda\neq0)$) and torsionless ones ($\tilde{T}_{fr}\equiv T_{fr}(\lambda=0)$). In particular, we find that
\begin{equation}
{\frac{\tilde{T}_{fr}}{T_{fr}}}= (1+2\lambda)^{1/3}\,.  \label{Tfr}
\end{equation}
Therefore, when $\lambda>0$ the effect of torsion is to \textit{reduce} the freeze-out temperature. As a result, the neutron-to-proton ratio ($\mathcal{N}=n/p$) will freeze-in at \textit{lower} temperatures and the residual helium-4 abundance will \textit{decrease} compared to that in the standard (torsion-free) Friedmann universe. The slowing of the expansion rate allows the neutron and protons to remain in non-relativistic kinetic equilibrium for longer, down to a lower temperature, with correspondingly fewer neutrons per proton surviving before the equilibrium is broken at $T_{fr}$.

This is a very rare (if not unique) example of a modified early-universe model with a reduced helium-4 abundance. All other common modifications (i.e.~extra light neutrino species, magnetic fields, anisotropies, Brans-Dicke fields, etc) lead to higher freeze-out temperatures. This increases the frozen-in $n/p$ ratio and therefore enhances the residual abundance of helium-4. In the presence of torsion this happens when $\lambda<0$.

We may quantify the above arguments by recalling that the frozen-in neutron-to-proton ratio is $\mathcal{N}_{fr}=\exp(-\Delta/T_{fr})$, where $\Delta$ measures the neutron-proton mass difference, and we have also set Boltzmann's constant to unity. Then, following (\ref{Tfr}), we arrive at the following simple relation
\begin{equation}
\mathcal{N}_{fr}= \tilde{\mathcal{N}}_{fr}^{\,(1+2\lambda)^{1/3}}\,, \label{cNs}
\end{equation}
between the torsional and the torsionless freeze-in ratios. Note that in the standard torsion-free early universe, $\tilde{\mathcal{N}}\simeq1/5$ at freeze-out, but falls to $\tilde{\mathcal{N}}\simeq1/7$ at the time of nucleosynthesis due to free neutron beta decay. In a universe without torsion, this means that the residual mass fraction of the synthesised helium-4 is $\tilde{Y}= 2\tilde{\mathcal{N}}_{ns}/(1+\tilde{\mathcal{N}}_{ns})\simeq0.25$. Assuming \textquotedblleft weak\textquotedblright torsion with $|\lambda|=|\phi|/H<1$ and keeping in mind that $\tilde{\mathcal{N}}_{ns}$, $\mathcal{N}_{ns}<1$, we obtain
\begin{equation}
Y= {\frac{2\mathcal{N}_{ns}}{1+\mathcal{N}_{ns}}}\simeq \mathcal{N}_{ns}\simeq \tilde{\mathcal{N}}_{ns}^{\,1+2\lambda/3}= \tilde{Y}\tilde{\mathcal{N}}_{ns}^{\,2\lambda/3}\simeq 0.25\times 7^{\,-2\lambda /3}\,.  \label{Y}
\end{equation}
Recent observational evidence for the allowed range of the primordial helium-4 abundance extrapolated to zero metals yields the range $0.2409\lesssim Y\lesssim0.2489$. Now using the standard prediction with zero torsion of $Y\simeq0.24703$~\cite{CFOY} with the observed range in Eq.~(\ref{Y}), lead to a strong observational constraint on the torsion parameter:
\begin{equation}
-0.005813\lesssim \lambda \lesssim +0.019370\,,  \label{lamcon}
\end{equation}
which is consistent with our $|\phi|/H<1$ assumption.

\section{Static spacetimes with torsion}\label{sSST} 
The extra degrees of freedom that torsion introduces are expected to relax some of the standard constraints associated with static spacetimes. We will therefore next turn our attention to the study of static (homogeneous and isotropic) models with torsion.

\subsection{The Einstein-static analogue}\label{ssE-SA} 
Static spacetimes with non-zero torsion have been studied in the past, assuming matter in the form of the Weyssenhoff fluid~\cite{W}. The latter, however, is incompatible with the high symmetry of the Friedmann-like models and thus with the Einstein static universe as well. For this reason, an unpolarised spin field was adopted, with a spin tensor that averages to zero (e.g.~see~\cite{Ku}). Here, instead, we address the FRW-compatibility issue by adopting a form for the torsion/spin fields that is compatible with the spatial isotropy and homogeneity of the Friedmannian spacetimes (see (\ref{ans.ten}) in \S~\ref{ssTA} earlier).

In static environments and in the absence of evolution, we may set $\dot{a}=0=\ddot{a}$ and $\dot{\rho}=0=\dot{p}= \dot{\phi}$. Then, the Friedmann equations derived in the previous section assume the (static) form
\begin{equation}
{\frac{1}{3}}\,\kappa\rho_{0}- {\frac{K}{a_{0}^{2}}}+ {\frac{1}{3}}\,\Lambda- 4\phi_{0}^{2}=0 \hspace{10mm} \mathrm{and} \hspace{10mm} {\frac{1}{2}}\,\kappa\left(\rho_{0}+3p_{0}\right)- \Lambda= 0\,,  \label{EScon}
\end{equation}
where $\rho_{0}$, $p_{0}$ and $\phi_{0}$ are constants. It follows that, when dealing with ordinary matter (i.e.~for $\rho_{0}>0$ and $\rho_{0}+3p_{0}>0$), the static solution requires the presence of a positive cosmological constant, just like the conventional Einstein-static universe (see Eq.~(\ref{EScon}b)). Unlike its Riemannian counterpart, however, the torsional analogue of the Einstein-static model does not necessarily require positive spatial curvature. Indeed, expression (\ref{EScon}a) guarantees that torsion can play the role of the positive curvature. Moreover, when $\phi_{0}\neq0$, the 3-curvature index can take all the available values (i.e.~$K=0,\pm1$). Also note that for matter with vanishing total gravitational energy, namely when $\rho_{0}+3p_{0}=0$, there is a static solution with Euclidean spatial hypersurfaces, zero cosmological constant, but non-zero torsion (i.e.~$K=0=\Lambda$ and $\rho_{0},\,\phi_{0}\neq0$).

Additional constraints come after successively eliminating the cosmological constant and the matter density from the set of (\ref{EScon}). In particular, we arrive at the following expressions
\begin{equation}
{\frac{1}{2}}\,\kappa\rho_{0}(1+w)- {\frac{K}{a_{0}^{2}}}= 4\phi_{0}^{2} \hspace{10mm} \mathrm{and} \hspace{10mm} {\frac{(1+w)\Lambda}{1+3w}}- {\frac{K}{a_{0}^{2}}}= 4\phi_{0}^{2}\,,  \label{EScon3}
\end{equation}
between the variables of the static model (with $w=p_{0}/\rho_{0}$ representing the barotropic index of the matter). Note that the last two constraints combine to reproduce (\ref{EScon}b). Finally, when $K=+1$, condition (\ref{EScon3}a) gives
\begin{equation}
a_0= \sqrt{{\frac{2}{\kappa\rho_0(1+w)-8\phi_0^2}}}\,, \label{ESradius}
\end{equation}
with $\kappa\rho_0(1+w)>8\phi_0^2$. Therefore, keeping the energy density of the matter fixed, the introduction of torsion increases the radius of the Einstein-static universe. Put another way, a torsional Einstein universe should be larger in size than its classic counterpart. A closely analogous effect, though in that case expressed in terms of the spin, was observed in~\cite{Bo}.

\subsection{Stability of the static model}\label{ssSSM} 
We will now test the stability of the static model, referring the reader to~\cite{BEMT} (and references therein) for analogous studies of the stability of the standard Einstein-static universe in general relativity. Here, we consider the evolution of linear conformal perturbations (i.e.~into other FRW models so that $\phi$ is not inhomogeneously perturbed) around the \textquotedblleft background\textquotedblright\ static solution of the previous section, In particular, assuming that $\phi=\phi_{0}=$~constant at the linear level and at all times, we consider small deviations of the form
\begin{equation}
a= a_{0}+ \delta a\,, \hspace{10mm} \rho= \rho_{0}+ \delta\rho \hspace{5mm} \mathrm{and} \hspace{5mm} p= p_{0}+ \delta p\,,  \label{lperts}
\end{equation}
where $\delta a\ll a_{0}$, $\delta\rho\ll\rho_{0}$ and $\delta p\ll p_{0}$ by construction. Substituting into Eqs.~(\ref{Fried1}) and (\ref{Fried2}), we solve (\ref{Fried1}) for the matter density. Then, using the resulting expression back in (\ref{Fried2}), employing the background relations (\ref{EScon})-(\ref{EScon3}), setting $\delta=\delta a/a_{0}\ll1$, and keeping up to first-order terms, we obtain the following differential equation,
\begin{equation}
a_{0}^{2}\,\ddot{\delta}+ 2(2+3w)\phi_{0}a_{0}^{2}\,\dot{\delta}- (1+3w)K\,\delta= 0\,,  \label{ddotdelta}
\end{equation}
for the linear evolution of the perturbation to the scale factor. Assuming matter with zero pressure and Euclidean spatial sections, namely setting $w=0=K$, the solution is
\begin{equation}
\delta= \mathcal{C}_{1}+ \mathcal{C}_{2}\mathrm{e}^{-4\phi_{0}t}\,,
\label{ES0}
\end{equation}
with $\mathcal{C}_{1,2}$ being the integration constants. This result shows (neutral) stability when $\phi_{0}>0$ and (exponential) instability for $\phi_{0}<0$. It is also straightforward to verify that Eq.~(\ref{ddotdelta}) leads to essentially the same solution, if the matter content satisfies the strong energy condition (i.e.~when $1+3w>0$). Therefore, an Einstein-static universe with non-zero torsion, conventional dust matter and flat spatial hypersurfaces can be stable when the associated (timelike) torsion vector is past-directed, but it is unstable when $S_{a}$ is future-directed (see (\ref{ans.vec}) in \S~\ref{ssTA}).

If we allow the 3-dimensional surfaces to have non-zero curvature, but maintain our assumption of pressureless matter (i.e.~$K=\pm1$ and $w=0$), Eq.~(\ref{ddotdelta}) solves to give
\begin{equation}
\delta=\mathcal{C}_{1}\mathrm{e}^{\alpha_{1}t}+ \mathcal{C}_{2}\mathrm{e}^{\alpha_{2}t}\,,  \label{ES1}
\end{equation}
where
\begin{equation}
\alpha_{1,2}= -2\phi_{0}\pm \sqrt{4\phi_{0}^{2}+{\frac{1}{a_{0}^{2}}}} \hspace{10mm} \mathrm{and} \hspace{10mm} \alpha_{1,2}= -2\phi_{0}\pm \sqrt{4\phi_{0}^{2}-{\frac{1}{a_{0}^{2}}}}\,,  \label{alphas}
\end{equation}
when $K=+1$ and $K=-1$ respectively. In the former case, solution (\ref{ES1}), (\ref{alphas}a) contains at least one (exponentially) growing mode, regardless of the sign of $\phi_{0}$ (i.e.~of the orientation of the torsion vector). Note also that the nature of the solution does not change so long as $1+3w>0$. Therefore, an Einstein-static universe with non-zero torsion, conventional matter and positively curved spatial hypersurfaces is always unstable. In models with negative 3-curvature the evolution is more involved. Following (\ref{ES1}) and (\ref{alphas}b), for $\phi_{0}^{2}a_{0}^{2}>1/4$, we have stability when $\phi_{0}>0$ and
instability for $\phi_{0}<0$. When $\phi_{0}^{2}a_{0}^{2}<1/4$, on the other hand, the solution of Eq.~(\ref{ES1}) contains an imaginary part. This translates into an oscillation with amplitude $\delta\propto\mathrm{e}^{-2\phi_{0}t}$. As before, the nature of the solution does not change so long as $1+3w>0$. Hence, an Einstein-static universe with non-zero torsion, conventional matter and open spatial hypersurfaces can be stable provided that $\phi_{0}>0$, namely when the associated torsion vector is past-directed.\footnote{As mentioned at the start of the current section, our stability analysis assumes homogeneous linear perturbations, similar to those employed by Eddington in his classic study of Einstein's static world~containing dust~\cite{E}, which was extended to other equations of state subjected to conformal perturbations in~\cite{Ha}. This implies that the stable configurations reported here may prove unstable in a more rigorous
investigation, where inhomogeneous perturbations of all the three possible types (i.e.~scalar, vector and tensor) are accounted for (see~\cite{BEMT} for a related linear-stability analysis on the classic Einstein-static spacetime and~\cite{BY} for the stability against general Mixmaster spatially homogeneous modes). Stable static universes also exist in general relativity with ghost fields ($\rho<0$), where stable bounded scale-factor oscillations of any amplitude occur around the static state~\cite{BT}.} An alternative method of testing the linear stability of the static solution, which arrives at the same conclusions but provides a different view point of the issue, is given in Appendix~C.

\section{Discussion}\label{sD}
The Einstein-Cartan gravity, as the simplest viable generalisation of classical general relativity, has attracted continuous attentions, especially since its re-introduction by Kibble and Sciama in the early 1960s. Allowing for an asymmetric connection, the theory incorporates the effects of spacetime torsion, which can then couple to the intrinsic angular momentum of the matter. Over the last six decades the Einstein-Cartan gravity and its variations have been applied to an extensive variety of theoretical problems, ranging from singularity theorems and cosmology, to supergravity and quantum gravity (e.g.~see~\cite{BH} and references therein).

The introduction of spacetime torsion is generally incompatible with the high symmetry of the FRW cosmologies. Therefore, one needs to consider torsion fields that preserve both the homogeneity and the isotropy of the host universe. In the present study we have addressed this issue by adopting a specific form for the torsion tensor, which belongs to the class of the vectorial torsion fields and introduces one additional degree of freedom, monitored by a single scalar function of time. Nevertheless, even this rather restricted form of torsion was found capable of drastically altering
the standard evolution-profile of the classic FRW universes.

We started by showing that, within the limits of our adopted torsion field, the Ricci, the Einstein and the matter energy-momentum tensors maintained the symmetry of their general relativistic counterparts. Using both $1+3$~covariant and metric-based techniques, we then derived the associated continuity, Friedmann and Raychaudhuri equations. These allowed us to quantify the relative strength of the torsion effects by means of an associated $\Omega $-parameter and a set of dimensionless variables. A number of new possibilities emerged. We found that torsion can play the role of the spatial curvature and mimic the effects of the cosmological constant, depending on the specifics of the scenario in hand. The orientation of the torsion vector relative to the fundamental 4-velocity field was also a decisive factor. Among others, it determines whether torsion will show a tendency to decelerate or accelerate the expansion of the host spacetime. The versatility of the torsion effects meant that empty spacetimes with zero 3-curvature and no cosmological constant are not necessarily static, but can experience exponential expansion (see also~\cite{ITP} for similar results). The introduction of spatial curvature, or matter, did not seem to change the aforementioned picture. We found that the torsion analogues of the Milne and the Einstein-de Sitter universes no longer exhibit their familiar coasting and power-law expansion, but can experience de Sitter-like inflation. All these findings raise the possibility that universes with non-zero torsion might have gone through an early (or a late) phase of accelerated expansion without requiring a cosmological constant, an inflaton field, or some sort of dark energy.

Looking for possible observational evidence of spacetime torsion, we
considered its effects on primordial nucleosynthesis. We found that torsion can increase, as well as decrease, the residual amount of helium-4, by changing the expansion rate of the universe at the time of primordial nucleosynthesis. Using our exact solution for the radiation-dominated Friedmann universe we were able to calculated the expected abundance of helium-4 from primordial nucleosynthesis when torsion is present. Combining these theoretical results with the current observationally allowed range for the helium-4 abundance, we were able to impose strong constraints on the strength of the associated torsion field.

Our study also concluded that there exist Einstein-static universes with torsion that are not necessarily closed, but can have all three types of spatial curvature. Unlike the classic (torsion-free) Einstein model, for appropriate choices of the torsion field and of the spatial curvature, these static universes can be stable against linear scalar perturbations even for pressureless (dust) matter. Overall, despite the restrictions imposed by their high symmetry, FRW-like universes with torsion exhibit a rich phenomenology, which could enable us to distinguish them from their standard general-relativistic counterparts.

\section{Appendices}
\appendix

\section{The generalised connection}\label{sAppA} 
According to the line element (\ref{FRWle}), the covariant metric tensor of a Friedmann-type spacetime (with or without torsion) has the diagonal form
\begin{equation}
g_{ab}=\left(\begin{matrix}
-1 & 0 & 0 & 0 \\ 0 & a^{2}/1-Kr^{2} & 0 & 0 \\ 0 & 0 & a^{2}r^{2} & 0 \\ 0 & 0 & 0 & a^{2}r^{2}\sin ^{2}\vartheta
\end{matrix}\right)\,,  \label{FRWmetrs}
\end{equation}
while its contravariant counterpart is $g^{ab}= \mathrm{diag}[-1,\,(1-Kr^{2})/a^{2},\,1/a^{2}r^{2},\, 1/a^{2}r^{2}\sin^{2}\vartheta]$. Therefore, the associated Christoffel symbols are (e.g.~see~\cite{Na})%
\footnote{The indices 0,1,2 and 3 correspond to the coordinates $t,r,\vartheta$ and $\varphi$ respectively.}
\begin{align}
& \tilde{\Gamma}^{0}{}_{11}= \frac{a\dot{a}}{1-Kr^{2}}\,,\quad \tilde{\Gamma}^{0}{}_{22}= a\dot{a}r^{2}\,,\quad \tilde{\Gamma}^{0}{}_{33}= a\dot{a}r^{2}\sin^{2}\vartheta\,,  \notag\\ & \tilde{\Gamma}^{1}{}_{01}= \tilde{\Gamma}^{1}{}_{10}= \frac{\dot{a}}{a}\,,\quad \tilde{\Gamma}^{1}{}_{11}= \frac{Kr}{1-Kr^{2}}\,,\quad \tilde{\Gamma}^{1}{}_{22}= -r\left(1-Kr^{2}\right)\,,\quad \tilde{\Gamma}^{1}{}_{33}=
-r\left(1-Kr^{2}\right)\sin^{2}\vartheta\,,  \notag\\
& \tilde{\Gamma}^{2}{}_{02}= \tilde{\Gamma}^{2}{}_{20}= \frac{\dot{a}}{a}\,,\quad \tilde{\Gamma}^{2}{}_{12}= \tilde{\Gamma}^{2}{}_{21}= \frac{1}{r}\,,\quad \tilde{\Gamma}^{2}{}_{33}= -\cos\vartheta\sin\vartheta\,,  \notag\\
& \tilde{\Gamma}^{3}{}_{03}= \tilde{\Gamma}^{3}{}_{30}= \frac{\dot{a}}{a}\,,\quad \tilde{\Gamma}^{3}{}_{13}= \tilde{\Gamma}^{3}{}_{31}= \frac{1}{r}\,,\quad \tilde{\Gamma}^{3}{}_{23}= \tilde{\Gamma}^{3}{}_{32}=
\cot\vartheta\,.  \label{Christ}
\end{align}
Combining (\ref{ans.ten}) with the above given covariant and contravariant forms of the metric tensor leads to the non-zero components of the torsion tensor, namely
\begin{equation}
S^{1}{}_{01}= S^{2}{}_{02}= S^{3}{}_{03}= \phi \hspace{10mm} \mathrm{and} \hspace{10mm} S^{1}{}_{10}= S^{2}{}_{20}= S^{3}{}_{30}= -\phi\,.  \label{torscomp}
\end{equation}
Putting (\ref{Christ}) and (\ref{torscomp}) together provides the components of the symmetric part of the generalised affine connection. In particular, recalling that $\Gamma^{a}{}_{(bc)}= \tilde{\Gamma}^{a}{}_{bc}+2S_{(bc)}{}^{a} $ -- see \S~\ref{ssTC} -- we find
\begin{align}
& \Gamma_{(11)}^{0}= \frac{a\dot{a}+ 2\phi a^{2}}{1-Kr^{2}}\,,\quad
\Gamma_{(22)}^{0}= r^{2}\left(a\dot{a}+2\phi a^{2}\right)\,,\quad
\Gamma_{(33)}^{0}= r^{2}\sin^{2}\vartheta\left(a\dot{a}+2\phi a^{2}\right)\,, \notag\\ & \Gamma_{(01)}^{1}= \frac{\dot{a}}{a}+\phi\,,\quad \Gamma_{(11)}^{1}= \frac{Kr}{1-Kr^{2}}\,,\quad \Gamma_{(22)}^{1}= -r\left(1-Kr^{2}\right)\,,\quad \Gamma_{(33)}^{1}= -r\left(1-Kr^{2}\right)\sin^{2}\vartheta\,,  \notag\\ & \Gamma_{(02)}^{2}= \frac{\dot{a}}{a}+ \phi\,,\quad \Gamma_{(12)}^{2}= \frac{1}{r}\,,\quad \Gamma_{(33)}^{2}= -\cos\vartheta\sin\vartheta\,, \notag\\
& \Gamma_{(03)}^{3}= \frac{\dot{a}}{a}+ \phi\,,\quad \Gamma_{(13)}^{3}= \frac{1}{r}\,,\quad \Gamma_{(23)}^{3}= \cot\vartheta\,.  \label{symGammas}
\end{align}
Finally, noting that $\Gamma^{a}{}_{bc}=\Gamma^{a}{}_{(bc)}+ S^{a}{}_{bc}$ and using the auxiliary results (\ref{torscomp}) and (\ref{symGammas}), we evaluate the non-zero components of the asymmetric affine connection
\begin{align}
& \Gamma^{0}{}_{11}= \frac{a\dot{a}+ 2\phi a^{2}}{1-Kr^{2}}\,,\quad
\Gamma^{0}{}_{22}= r^{2}\left(a\dot{a}+2\phi a^{2}\right)\,,\quad
\Gamma^{0}{}_{33}= r^{2}\sin^{2}\vartheta\left( a\dot{a}+2\phi
a^{2}\right)\,,  \notag \\ & \Gamma^{1}{}_{01}= \frac{\dot{a}}{a}+ 2\phi\,,\quad \Gamma^{1}{}_{10}= \frac{\dot{a}}{a}\,,\quad \Gamma^{1}{}_{11}= \frac{Kr}{1-Kr^{2}}\,,  \notag\\ & \Gamma^{1}{}_{22}= -r\left(1-Kr^{2}\right)\,,\quad \Gamma^{1}{}_{33}= -r\left(1-Kr^{2}\right)\sin^{2}\vartheta\,,  \notag\\ & \Gamma^{2}{}_{02}= \frac{\dot{a}}{a}+ 2\phi\,,\quad \Gamma^{2}{}_{20}= \frac{\dot{a}}{a}\,,\quad \Gamma^{2}{}_{12}= \Gamma^{2}{}_{21}= \frac{1}{r}\,,\quad \Gamma^{2}{}_{33}= -\cos\vartheta\sin\vartheta\,, \notag\\ & \Gamma^{3}{}_{03}= \frac{\dot{a}}{a}+ 2\phi\,,\quad \Gamma^{3}{}_{30}= \frac{\dot{a}}{a}\,,\quad \Gamma^{3}{}_{13}= \Gamma^{3}{}_{31}= \frac{1}{r}\,,\quad \Gamma^{3}{}_{23}= \Gamma^{3}{}_{32}= \cot\vartheta\,.  \label{Gammas}
\end{align}

\section{Ricci tensor and Friedmann equations}\label{sAppB} 
The Ricci curvature tensor has been expressed in terms of the generalised (asymmetric) affine connection in \S ~\ref{ssFEs} -- see Eq.~(\ref{Ricci}) there. The latter, together with (\ref{Gammas}), leads to
\begin{align}
& R_{00}= -3\left[{\frac{\ddot{a}}{a}}+2\dot{\phi} +2\phi\left({\frac{\dot{a}}{a}}\right)\right]\,, \notag\\
& R_{11}= \frac{1}{1-Kr^{2}}\left(a\ddot{a}+2\dot{\phi}a^{2} +2\dot{a}^{2}+10\phi a\dot{a}+8\phi^{2}a^{2}+2K\right)\,, \notag\\
& R_{22}= r^{2}\left(a\ddot{a}+2\dot{\phi}a^{2}+2\dot{a}^{2}+10\phi a\dot{a}+8\phi^{2}a^{2}+2K\right)\,, \notag\\ & R_{33}= r^{2}\sin^{2}\vartheta\left(a\ddot{a} +2\dot{\phi}a^{2}+2\dot{a}^{2}+10\phi a\dot{a}+8\phi^{2}a^{2} +2K\right)\,.  \label{covRicci}
\end{align}
Given that $R_{a}{}^{b}=g^{bc}R_{ac}$, the above list combines with the contravariant form of the FRW metric (see Appendix~A above) to provide the mixed components of the Ricci tensor, namely
\begin{equation}
R_{0}{}^{0}= 3\left[\frac{\ddot{a}}{a}+2\dot{\phi} +2\phi\left(\frac{\dot{a}}{a}\right)\right]  \label{mixRicci1}
\end{equation}
and
\begin{equation}
R_{1}{}^{1}= R_{2}{}^{2}= R_{3}{}^{3}= \left(a\ddot{a}+2\dot{\phi}a^{2}+2\dot{a}^{2}+10\phi a\dot{a} +8\phi^{2}a^{2}+2K\right)\,.  \label{mixRicci2}
\end{equation}
Consequently, the Ricci scalar of an FRW-like spacetime with non-zero torsion reads
\begin{equation}
R= 6\left[\frac{\ddot{a}}{a}+\left(\frac{\dot{a}}{a}\right)^{2} +\frac{K}{a^{2}}+2\dot{\phi}+6\phi\left(\frac{\dot{a}}{a}\right) +4\phi^{2}\right]\,.  \label{scRicci}
\end{equation}
Finally, we introduce the energy-momentum tensor of a perfect fluid, which takes the diagonal form $T_{a}{}^{b}= \mathrm{diag}[-\rho,p,p,p]$. Then, plugging all of the above into the Einstein-Cartan field equations (see expression (\ref{E-CFEs}) in \S~\ref{ssFEBIs}), we arrive at
\begin{equation}
\left(\frac{\dot{a}}{a}\right)^{2}= {\frac{1}{3}}\,\kappa\rho- \frac{K}{a^{2}}+ {\frac{1}{3}}\,\Lambda- 4\phi\left(\frac{\dot{a}}{a}\right)- 4\phi^{2}\,,
\end{equation}
and
\begin{equation}
\frac{\ddot{a}}{a}= -{\frac{1}{6}}\,\kappa\left(\rho+3p\right)+ {\frac{1}{3}}\,\Lambda- 2\dot{\phi}- 2\phi\left(\frac{\dot{a}}{a}\right)\,,
\end{equation}
which are the Friedmann equations of an FRW-type cosmology with torsion (see \S~\ref{ssFEs} earlier).

\section{Linear stability of the static solution}\label{sAppC}
Here, to complement the analysis given in \S ~\ref{sssESM}, we will employ the linear stability technique of ordinary differential equations to test the stability of the static model. In doing so, we first recast Eq.~(\ref{ddotdelta}) into the following system of first-order differential equations
\begin{equation}
\dot{x}= f(x,y)=y \hspace{10mm} \mathrm{and} \hspace{10mm} \dot{y}= g(x,y)= \mu y+ \nu x\,.  \label{system}
\end{equation}
with $x=\delta$, $\mu=-2(2+3w)\phi_{0}$ and $\nu=
\left(1+3w\right)K/a_{0}^{2}$. The Jacobian matrix of the above given linearized system is
\begin{equation}
\left(\begin{matrix}
\partial f/\partial x & \partial f/\partial y \\ & \\
\partial g/\partial x & \partial g/\partial y%
\end{matrix}\right)_{0}=
\left(\begin{matrix}
0 & 1 \\ \nu & \mu
\end{matrix}\right)\,,  \label{Jacobian}
\end{equation}
where the zero suffix denotes the static solution (with $(x=0,y=0$) and $f(0,0)=0=g(0,0)$). The eigenvalues of matrix (\ref{Jacobian}) are the roots of the associated characteristic polynomial, namely
\begin{equation}
\lambda_{1,2}= {\frac{1}{2}} \left(\mu\pm\sqrt{\mu^{2}+4\nu}\right)\,.  \label{lambdas1}
\end{equation}
For the sake of simplicity, though without compromising generality, let us confine to the case of pressure-free matter (with $w=0$ -- see also \S~\ref{ssSSM} earlier). Then, the eigenvalues reduce to
\begin{equation}
\lambda _{1,2}= -2\phi_{0}\pm \sqrt{4\phi_{0}^{2}+{\frac{K}{a_{0}^{2}}}}\,.
\end{equation}
When the spatial hypersurfaces have spherical geometry (i.e.~for $K=+1$), the quantity inside the square root is always positive, in which case both eigenvalues are real. More specifically, $\lambda_{1}$ is positive and $\lambda_{2}$ is negative, implying that the static solution is a saddle point. For hyperbolic spatial surfaces (i.e.~when $K=-1$), the two eigenvalues are real provided that $a_{0}^{2}\phi_{0}^{2}>1/4$. Then, both $\lambda_{1}$ and $\lambda_{2}$ are positive when $\phi_{0}<0$, while for $\phi_{0}>0$ the two eigenvalues are negative. In the former case the static solution is unstable, but in the latter is stable. Alternatively, when $a_{0}^{2}\phi_{0}^{2}<1/4$, the eigenvalues are both complex. In particular, $Re(\lambda_{1})>0$ and $Re(\lambda_{2})>0$ for $\phi_{0}<0$, which implies instability. For $\phi_{0}>0$, on the other hand, we have $Re(\lambda_{1,2})<0$ and stability. The same is also true when $a_{0}^{2}\phi_{0}^{2}=1/4$.

Finally, note that the above analysis does not apply to spacetimes with Euclidean spatial sections (i.e.~when $K=0$), because then the first of the two eigenvalues\ will vanish. In such a case, in order to study the stability of the static solution, one needs to solve Eq.~(\ref{ddotdelta}) analytically (see \S~\ref{ssSSM} previously).\newline

\textbf{Acknowledgements:} We would like to thank Peter Stichel for drawing our attention to reference~\cite{T}. CGT acknowledges support from a visiting fellowship by Clare Hall College and visitor support by DAMTP at the University of Cambridge, where part of this work took place. JDB was supported by the Science and Technology Facilities Council (STFC) of the UK.


\begin{thebibliography}{99}
\bibitem{C} E. Cartan, C.R. Acad. Sci. (Paris) 174, 593 (1922); E. Cartan, Ann. Sci. Ec. Norm. Super. \textbf{40}, 325 (1923); \textit{ibid}, \textbf{41}, 1 (1924); \textit{ibid}, \textbf{42}, 17 (1925).
\bibitem{K} T.W.B. Kibble, J. Math. Phys. \textbf{2}, 12 (1961); D. Sciama, Rev. Mod. Phys. \textbf{36}, 463 (1964).
\bibitem{BH} M. Blagojevic and F.W. Hehl, \textit{Gauge Theories and Gravitation} (Imperial College Press, London, 2013).
\bibitem{H} R.T. Hammond, Rept. Prog. Phys. \textbf{65}, 599 (2002); Y. Mao, M. Tegmark, A.H. Guth and Cabi S., Phys. Rev. D \textbf{76}, 104029 (2007); V.A. Kostelecky, Russell N. and Tasson J., Phys. Rev. Lett. \textbf{100}, 111102 (2008); R. March, G. Bellettini, R. Tauraso and S. Dell'Angello, Phys. Rev. D \textbf{83}, 104008 (2011); F.W. Hehl, Y.N. Obukhov and D. Puetzfeld, Phys. Lett. A \textbf{377}, 1775 (2013); D. Puetzfeld and Y.N. Obukhov, Int. J. Mod. Phys. D \textbf{23}, 1442004 (2014); R.-H. Lin, X.H. Zhai and X.Z. Li, Eur. Phys. J. C \textbf{77}, 504 (2017).
\bibitem{CLS} S. Capozziello, G. Lambiase and C. Stornaiolo, Annalen Phys. \textbf{10}, 713 (2001).
\bibitem{T} M. Tsamparlis, Phys. Rev. D \textbf{24}, 1451 (1981).
\bibitem{O} G.J. Olmo, Int. J. Mod. Phys. D \textbf{20}, 413 (2011); S. Capozzielo, R. Cianci, C. Stornaiolo and S. Vignolo, Phys. Scripta \textbf{78}, 065010 (2008); Beltran Jimenez J. and Koivisto T.S., Phys. Lett. B \textbf{756}, 400 (2016).
\bibitem{P} N.J. Poplawski, Phys. Lett. B \textbf{694}, 181 (2010); N.J. Poplawski, Astron. Rev. \textbf{8}, 108 (2013); A.N. Ivanov and M. Wellenzohn, Astrophys. J. \textbf{829}, 47 (2016); S. Akhshabi, E. Qorani and F. Khajenabi, Europhys. Lett. \textbf{119}, 29002 (2017); R. Banerjee, S. Chakraborty and P. Mukherjee, Phys. Rev. D \textbf{98}, 083506 (2018).
\bibitem{S} L.L. Smalley, Phys. Rev. D \textbf{18}, 3896 (1978).
\bibitem{PTB} K. Pasmatsiou, C.G. Tsagas and J.D. Barrow, Phys. Rev. D \textbf{95}, 104007 (2017).
\bibitem{HvdHK} F.W. Hehl, P. von der Heyde and G.D. Kerlick, Rev. Mod. Phys. \textbf{48}, 373 (1976); F.W. Hehl and Y.N. Obukhov, Annales Fond. Broglie \textbf{32}, 157 (2007).
\bibitem{TCM} C.G. Tsagas, A. Challinor and R. Maartens, Phys. Rep. \textbf{465}, 61 (2008); G.F.R. Ellis, R. Maartens and M.A.H. MacCallum, \textit{Relativistic Cosmology} (Cambridge University Press, Cambridge, 2012).
\bibitem{N} H. Nariai, Prog. Theor. Phys. \textbf{40}, 48 (1969); C. Mathiazhagen and V.B. Johri, Class. Quantum Grav. \textbf{1}, L29 (1984); J.D. Barrow and K-I. Maeda, Nucl. Phys. B\textbf{341}, 294 (1990).
\bibitem{B} J.D. Barrow, Phys. Lett. B \textbf{235}, 40 (1990).
\bibitem{CFOY} R.H. Cyburt, B.D. Fields, K.A. Olive, and T.-H. Yeh, Rev.Mod. Phys. \textbf{88}, 015004 (2016).
\bibitem{W} J. Weyssenhoff and J. Raade, Acta Phys. Pol. \textbf{9}, 7 (1947).
\bibitem{Ku} B. Kuchowicz, Gen. Rel Grav. \textbf{9}, 511 (1978); M. Gasperini, Phys. Rev. Lett. \textbf{56}, 2873 (1986); K. Atazadeh, JCAP \textbf{06}, 020 (2014).
\bibitem{Bo} C.G. B\"ohmer, Class. Quantum Grav. \textbf{21}, 1119 (2004).
\bibitem{BEMT} J.D. Barrow, G.F.R. Ellis, R. Maartens and C.G. Tsagas, Class. Quantum Grav. \textbf{20}, L155 (2003).
\bibitem{E} A.S. Eddington, Mon. Not. R. Astron. Soc. \textbf{90}, 668 (1930).
\bibitem{Ha} E.R. Harrison, Rev. Mod. Phys. \textbf{39}, 862 (1967); G.W. Gibbons, Nucl. Phys. B 292, 784 (1987); \textit{ibid}, \textbf{310}, 636 (1988).
\bibitem{BY} J.D. Barrow and K. Yamamoto, Phys. Rev. D \textbf{85}, 083505 (2012).
\bibitem{BT} J.D. Barrow and C.G. Tsagas, Class. Quantum Grav. \textbf{26}, 195003 (2009); J.D. Barrow, D. Kimberly and J. Magueijo, Class. Quantum Grav. \textbf{\ 21}, 4289 (2004).
\bibitem{ITP} D. Iosifidis, C.G. Tsagas and A.C. Petkou [arXiv:1809.04992].
\bibitem{Na} J.V. Narlikar, \textit{Introduction to Cosmology} (Jones and Bartlett, Boston, 1983).
\end{thebibliography}
\end{document}